\documentclass[journal]{IEEEtran}

\usepackage{cite}

\ifCLASSINFOpdf
   \usepackage[pdftex]{graphicx}
   \graphicspath{{../pdf/}{../jpeg/}}
   \DeclareGraphicsExtensions{.pdf,.jpeg,.png}
\else
\fi
\UseRawInputEncoding
\usepackage{amsmath}
\usepackage{array}
\usepackage{standalone}
\usepackage{etoolbox}
\newtoggle{arXiv}
\newtoggle{bibrun}
\usepackage{newtxtext, newtxmath}
\usepackage{graphicx}
\usepackage{color, calc}
\usepackage{array, booktabs}
\newcolumntype{L}{>{$}l<{$}}
\newcolumntype{C}{>{$}c<{$}}
\newcolumntype{R}{>{$}r<{$}}
\usepackage{tikz, tikz-3dplot}
\usetikzlibrary{arrows, arrows.meta, calc, positioning}
\usetikzlibrary{decorations.pathmorphing}	
\tikzset{>=Stealth}
\usepackage{siunitx, physics}
\usepackage{enumitem}
\setlist[description]{labelindent=0pt, leftmargin=\parindent, font=\normalfont\itshape}
\usepackage{url}
\usepackage{subfigure}
\usepackage{textcomp}
\usepackage{eurosym}
\usepackage{xfrac}
\usepackage{pgfplots}
\usepackage{stmaryrd}
\pgfplotsset{compat=1.17}
\usepackage{makecell}

\newcommand{\M}{\mbox{MaRCoS}}
\newcommand{\MG}{\mbox{MaRGA}}

\begin{document}
%
\title{\MG{}: a Graphical and Application Interface for \M{}}

\author{\IEEEauthorblockN{
		Jos\'e\,M.\,Algar\'{\i}n\IEEEauthorrefmark{1},
		Teresa\,Guallart-Naval\IEEEauthorrefmark{1},
		Jos\'e\,Borreguero\IEEEauthorrefmark{2},
		Fernando\,Galve\IEEEauthorrefmark{1}, and
		Joseba\,Alonso\IEEEauthorrefmark{1}}
	
	\IEEEauthorblockA{\IEEEauthorrefmark{1}MRILab, Institute for Molecular Imaging and Instrumentation (i3M), Spanish National Research Council (CSIC) and Universitat Polit\`ecnica de Val\`encia (UPV), 46022 Valencia, Spain}\\
	\IEEEauthorblockA{\IEEEauthorrefmark{2}Tesoro Imaging S.L., 46022 Valencia, Spain}\\
	
\thanks{Corresponding author: J. Alonso (joseba.alonso@i3m.upv.es).}}


\maketitle

\begin{abstract}
	The open-source console \M{}, which stands for ``Magnetic Resonance Control System'', combines hardware, firmware and software elements for integral control of Magnetic Resonance Imaging (MRI) scanners. Previous developments have focused on making the system robust and reliable, rather than on users, who have been somewhat overlooked. This work describes a Graphical User Interface (GUI) designed for intuitive control of \M{}, as well as compatibility with clinical environments. The GUI is based on an uncomplicated set of panels and a renewed Application Program Interface (API). Compared to the previous versions, the \MG{} package (``\M{} Graphical Application'') includes new functionalities such as the possibility to export images to standard DICOM formats, create and manage clinical protocols, or display and process image reconstructions, among other features conceived to simplify the operation of MRI scanners. All prototypes in our facilities are commanded by MaRCoS and operated with the new GUI. Here we report on its performance on an experimental 0.2~T scanner designed for hard-tissue imaging, as well as a 72~mT portable scanner presently installed in the radiology department of a large hospital. The possibility to customize, adapt and streamline processes has substantially improved our workflows and overall experience.
\end{abstract}


 \ifCLASSOPTIONpeerreview
 \begin{center} \bfseries EDICS Category: 3-BBND \end{center}
 \fi
%
\IEEEpeerreviewmaketitle


\section{Introduction}\label{sec:Intro}
Until recently, the availability of Magnetic Resonance Imaging (MRI) has been strongly compromised by the initial investment required, as well as maintenance and infrastructural costs. A transformative shift is taking place, however, driven by a collective endeavor to reduce expenses and enhance affordability. Particularly, the advent of low-field MRI technology has significantly contributed to pushing down costs, making MRI more accessible to a broader population, transcending economic barriers and ensuring that the benefits of this medical imaging modality reach diverse communities \cite{Sarracanie2020, Arnold2022, Webb2023}.

The electronic control system or `console' is pivotal to any MRI scanner. This component needs to control an intricate interplay of electromagnetic pulses with exquisite time resolution, as well as manage signal acquisition and the subsequent image reconstruction. Traditionally, MRI scanners have operated using proprietary consoles, but this constrains their adaptability and extensibility \cite{Stang2012}. Therefore, as low-field MRI systems gain momentum, the need for an open-source, adaptable, and affordable console continues to grow. Among the numerous open-source initiatives to develop an affordable and widely spread control system \cite{Tang2015, HASSELWANDER201647, Anand2018, OCRA, TAKEDA2011355, Michal2018, ARIANDO201974, ZHEN2021106852, Layton2017}, the recently released \emph{Magnetic Resonance Control System} (\M{}) stands out for its high performance, versatility and low cost \cite{GuallartNaval2022b, Negnevitsky2023}.

Understandably, the first developments of \M{} have focused on hardware and firmware, advancing towards a robust and reliable technology while keeping the costs down and promoting accessibility. Nevertheless, \M{} is a live project developed collaboratively within an open international community and designed for massive use in a wide variety of MRI systems, so the software needs to eventually transition from machine-centered to user-centered. We consider that the next milestone in this regard is the release of a Graphical User Interface (GUI) that facilitates the installation of \M{} in MRI systems, regardless of their hardware, and that eases scanner operation both in laboratory and clinical settings. In this vision, it is critical that the interaction between human operators and the scanner is intuitive, but also that the GUI offers an accessible gateway for users to modify sequence parameters, apply imaging protocols, and visualize results.

In previous publications we introduced a rudimentary GUI designed to offer fundamental functionality for any system controlled by \M{} \cite{GuallartNaval2022b, Negnevitsky2023}. However, this first version had a number of limitations:  i) communication management with the \M{} server required users to establish a connection with the server through terminal commands before accessing the GUI; ii) session management lacked information, leading to potential data mismanagement and disorganization; iii) the GUI only allowed single sequence runs, as opposed to the multi-sequence protocols which MRI technologists and radiologists are used to; iv) the absence of an automated calibration protocol for hardware components required users to manually calibrate the system parameters (e.g. Larmor frequency or gradient shimming) and introduce them in the next executed sequences, which was error-prone and severely limited the workflow; v) the GUI lacked functionalities required for integration in clinical environments, e.g. DICOM compatibility, making it clash with established workflows; vi) the inability to compare different reconstructions acquired during the same session hindered image analysis and diagnostic evaluations; and vii) the GUI did not include tools for image post-processing.

In this paper, we present \MG{}, a major upgrade with respect to the existing \M{} GUI, fully coded in Python~3, and designed to overcome the aforementioned limitations. These upgrades are aimed at ensuring the adaptability and usability of the system, both in controlled research environments and in clinical practice. The text describes technical details of \MG{} and provides insight on its underlying Application Program Interface (API). Finally, we present results obtained from experiments conducted in a laboratory device dedicated to dental imaging \cite{Algarin2020, Borreguero2023} and a portable unit for extremity imaging \cite{GuallartNaval2022, Algarin2023} installed in the Medical Imaging Department of La Fe Hospital in Valencia, Spain.

\section{Goals}\label{sec:Goals}

\begin{figure*}[t]
	\centering
	\includegraphics[width=\textwidth]{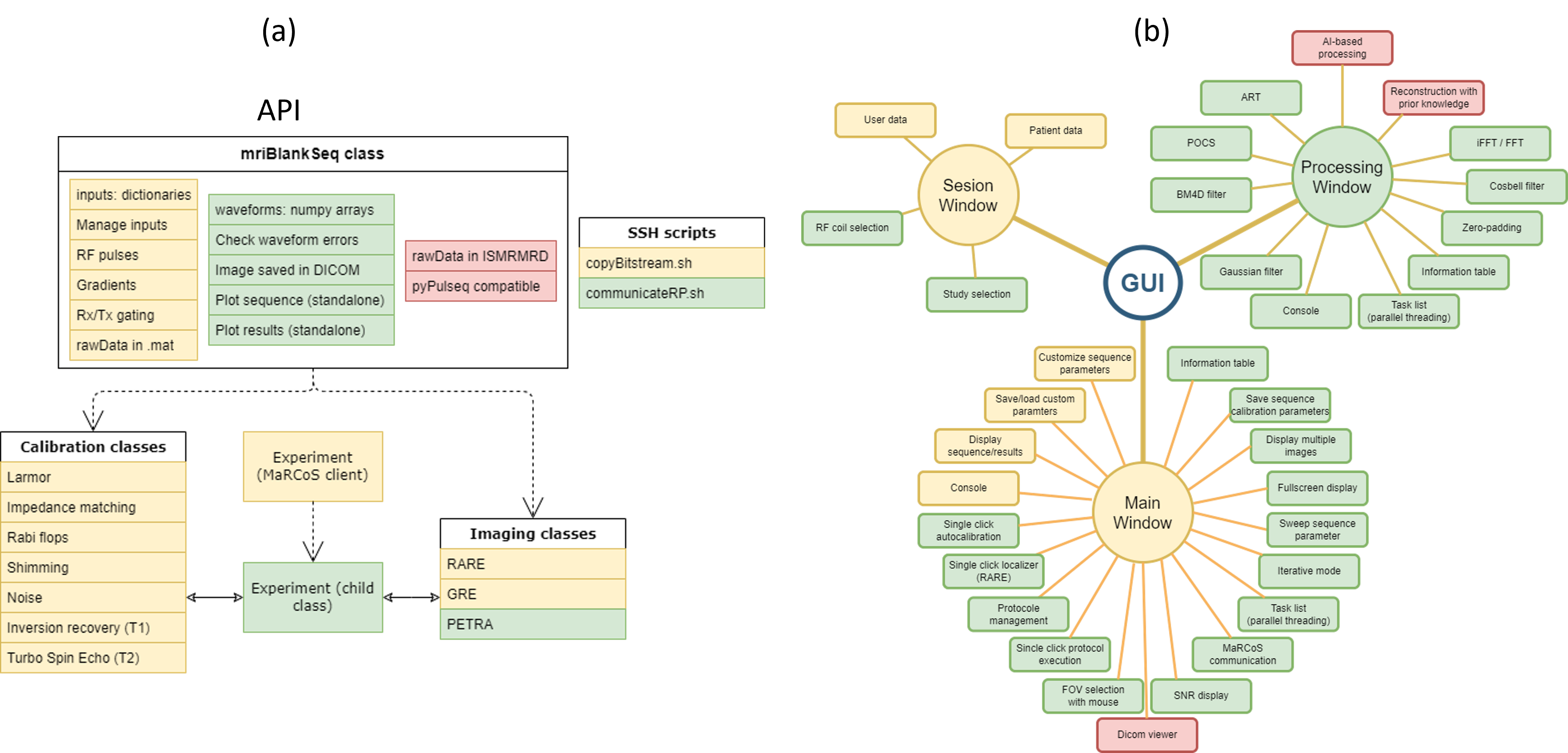}
	\caption{Diagram of the API (a) and GUI (b). Colors represent developments in the first release (orange), what has been implemented in current version (green) and possible additional implementations (red). For the API sketch, dashed (solid) arrows represents inheritance (communication). The \texttt{mriBlankSeq} class box contains some of its attributes and methods.}
	\label{fig:diagram}
\end{figure*}

The main goals behind \MG{} are three: first, to improve the user interaction with the \M{} system, fostering the experience for human operators with varying levels of technical expertise; second, to optimize the functionality of \M{} specifically for laboratory and medical environments, tailoring its capabilities for diverse research and clinical needs; and third, to provide a set of tools for image post-processing. To meet these objectives, we have tried to design the software placing the focus on users and their experience.

\section{Application Program Interface}\label{sec:API}

The \MG{} API (Fig.~\ref{fig:diagram}(a)) has been programmed in Python and encapsulates functionalities intended to simplify the interaction between users and the \M{} system, including establishing communication with the \M{} server, executing pulse sequences, or managing data storage.

\subsection{Setting up communication with \M{} server}

Upon initialization, the GUI needs to establish communication with the \M{} server running in the Linux environment on the SDRLab board \cite{Negnevitsky2023}. This process requires setting up the network parameters and ensuring the integrity of data transmission. Encrypted communication with the \M{} server is performed with two Secure Shell (SSH) files: \texttt{copyBitstream.sh} to setup MaRCoS into the SDRLab, and \texttt{communicateRP.sh} to open and close the bidirectional communication channels with the \M{} server. This communication allows users to send sequences from the client PC, controlling the system at a high level, or receiving data arrays back on the client PC once the sequence is finished \cite{Negnevitsky2023}.

\subsection{\texttt{mriBlankSeq}: the master class to create pulse sequences}

At the core of the API lies the \texttt{mriBlankSeq} class (Fig.~\ref{fig:diagram}(a)), an extensible parent class that serves as the base for generating arbitrary pulse sequences. In essence, this master class provides a structured framework for creating, executing, and managing complex MRI sequences. To that end, \texttt{mriBlankSeq} incorporates a number of attributes and methods (see mriBlankSeq box in Fig.~\ref{fig:diagram}(a)).

A first group of attributes include dictionaries to store sequence inputs and waveform data. They therefore function as containers for parameters that define the characteristics of an MRI sequence, including timing, pulse amplitudes and phases, gradient settings and others. Additionally, the \texttt{mriBlankSeq} class integrates methods for defining radiofrequency (RF) pulses, gradient waveforms, and acquisition triggers. Leveraging these methods, users can assemble complex sequences by concatenating smaller waveform components.

In the current version of the API, the complete generated waveform is stored in NumPy arrays, which are subsequently transmitted to the \M{} client in the control computer. The communication between the sequence generator and the \M{} server on the SDRLab makes use of a dedicated method within the master class (\texttt{floDict2Exp}), which checks for inconsistencies on the sequence waveforms and orchestrates the transfer of waveform data to the \M{} server. These checks guarantee that the external electronics receive the right analog pulses, e.g. by identifying negative time steps or checking pulse amplitudes are within hardware bounds.

In \MG{}, reconstructions are saved in the DICOM 3.0 standard format for medical image storage and interoperability, as well as a Matlab native file (\texttt{.mat}) including the raw signal data acquired by the scanner, the reconstructed images, and relevant metadata related to executed sequences. This information facilitates subsequent analyses, comparisons, and validations of sequence performance. The corresponding \texttt{mriBlankSeq} method is called \texttt{saveRawData}.

\begin{figure}[h]
	\centering
	\includegraphics[width=\columnwidth]{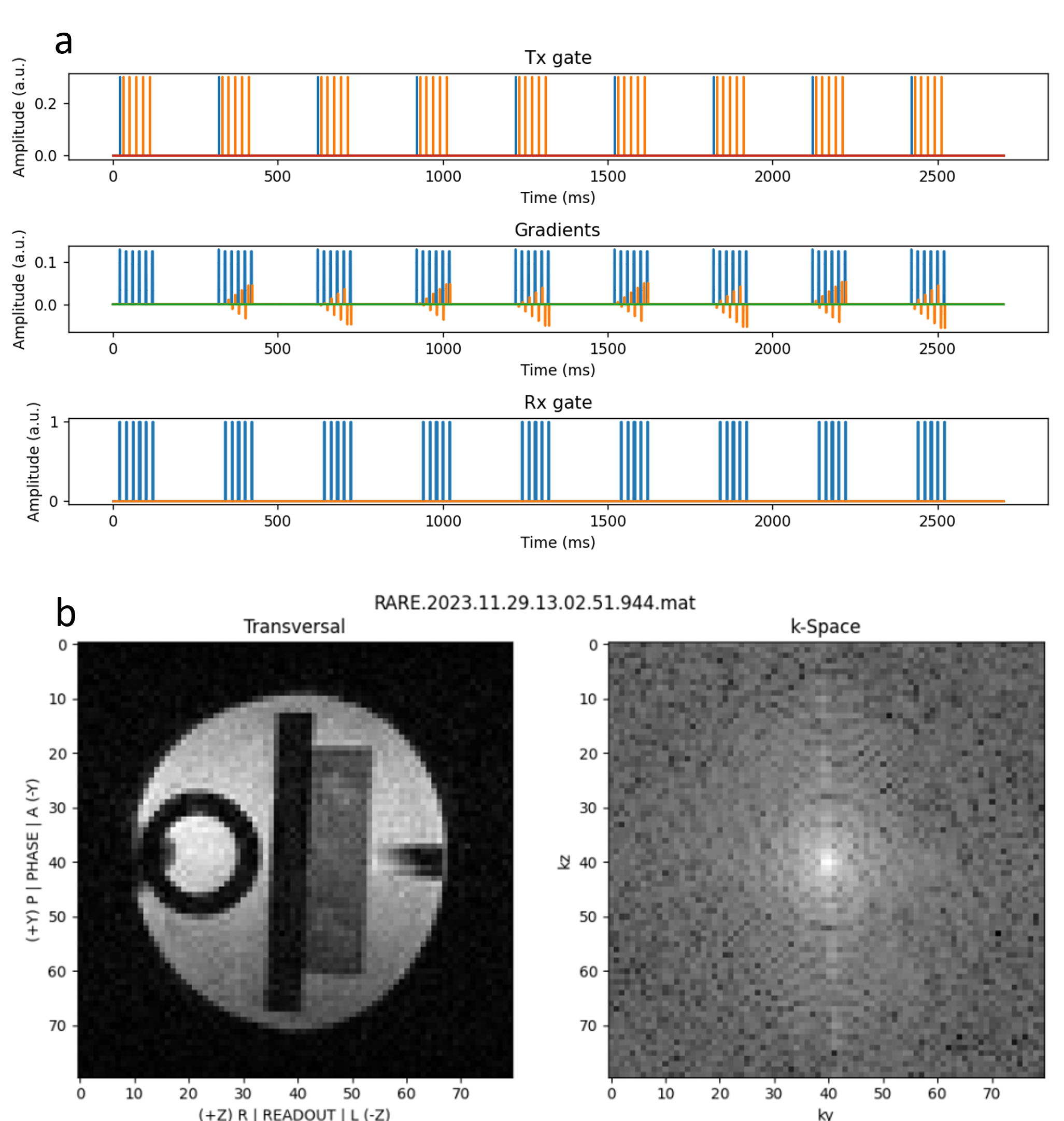}
	\caption{Screenshot of the results displayed after a sequence is executed in standalone mode. a) Sequence waveform display after calling the \texttt{sequencePlot} method. b) Image display after calling the \texttt{plotResults} method.}
	\label{fig:api_plot}
\end{figure}

\texttt{mriBlankSeq} further integrates two methods for data visualization: \texttt{sequencePlot} can be invoked to display the sequence instructions (Fig.~\ref{fig:api_plot}(a)), offering a graphical overview of the designed pulse patterns; whereas \texttt{plotResults} provides a visual representation of the resulting data (Fig.~\ref{fig:api_plot}(b)), which can take the form of one or two-dimensional plots depending on whether the user wants to visualize a signal or an image. These methods are called later by the GUI to visualize data (Figs.~\ref{fig:gui_screenshot}b, \ref{fig:DentalI} and \ref{fig:PhysioI}), but may be run also in standalone execution (i.e. without the GUI), which we find useful during debugging (Fig.~\ref{fig:api_plot}).

In conclusion, the \texttt{mriBlankSeq} class stands as the cornerstone of the API, encompassing user inputs, waveform generation, server communication, data preservation and visualization. Through its attributes and methods, \texttt{mriBlankSeq} allows users to generate complex MRI sequences and integrate them within the \M{} environment for execution and data management.

\subsection{Pulse sequences: child classes}

\begin{figure*}[t]
	\centering
	\includegraphics[width=0.7\textwidth]{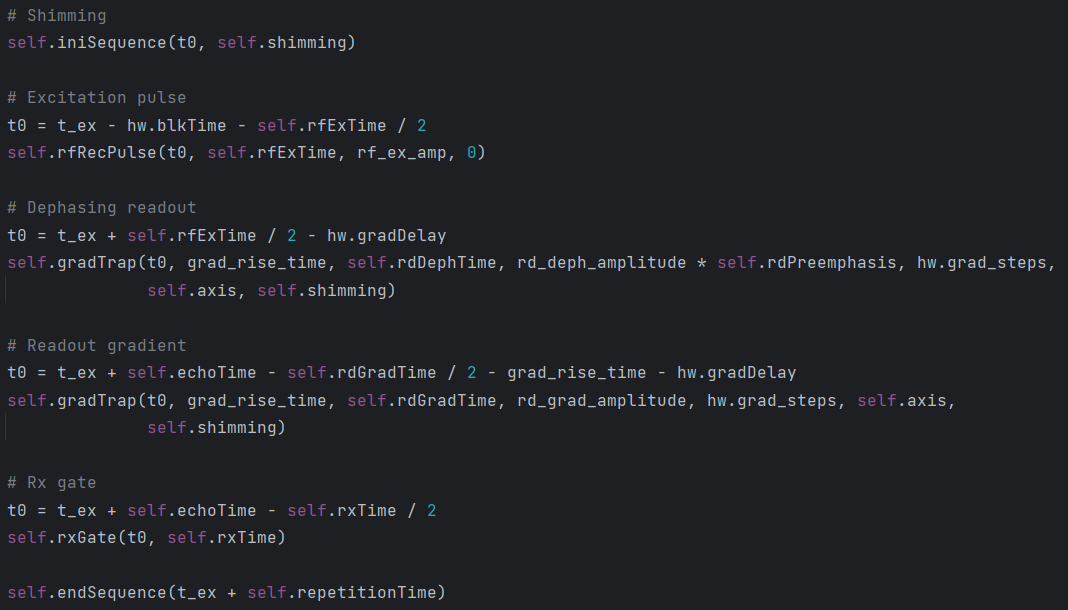}
	\caption{Sample code for a one-dimensional gradient echo pulse sequence in \MG{}.}
	\label{fig:sequence}
\end{figure*}

At the heart of the sequence hierarchy lies the principle of inheritance. In the \MG{} API, each pulse sequence class is conceived as a child of \texttt{mriBlankSeq}, thereby inheriting its attributes and methods. This design approach ensures a consistent and standardized foundation across all sequence classes, facilitating code management and ideally promoting best practices. By adopting this object-oriented inheritance model, \MG{} fosters modular development, allowing programmers to leverage the existing functionality provided by \texttt{mriBlankSeq} and build upon it to craft different sequences.

Child classes are executed by invoking the \texttt{sequenceRun} method. Figure~\ref{fig:sequence} showcases a simplified representation of the sequence structure for a 1D gradient echo. Once a pulse sequence is executed, the acquired data can be reconstructed into images for analysis and visualization by means of the \texttt{sequenceAnalysis} method. Sequences with a high number of acquired points can be executed in batches to prevent buffer overflow. The maximum allowable number of points before initiating a new batch can be configured through the hardware configuration file (Sec.~\ref{subsec:HardwareConfig}). 

\subsection{Hardware Configuration File}\label{subsec:HardwareConfig}

The hardware specifications of an MRI system determine to a large extent the quality and accuracy of the reconstructed images. The open, modular nature of \M{} means that the software needs adjustment whenever new hardware is encountered. To make this scalable, the \MG{} API includes a configuration file that stores parameters related to the scanner hardware components, acting as a centralized repository for information on RF coils (e.g. serial number), RF and gradient amplifiers (gain, input parameter ranges, etc.), or digital components such as the SDRLab boards or the control computer (IP addresses, etc.). This information is used by the API to initialize and interact with the MRI system's hardware, ensuring accurate control and synchronization during MRI scans. The hardware configuration file can be manually written following a simple structure \cite{MARGA}.

\section{Graphical User Interface}

\begin{figure*}[t]
	\centering
	\includegraphics[width=\textwidth]{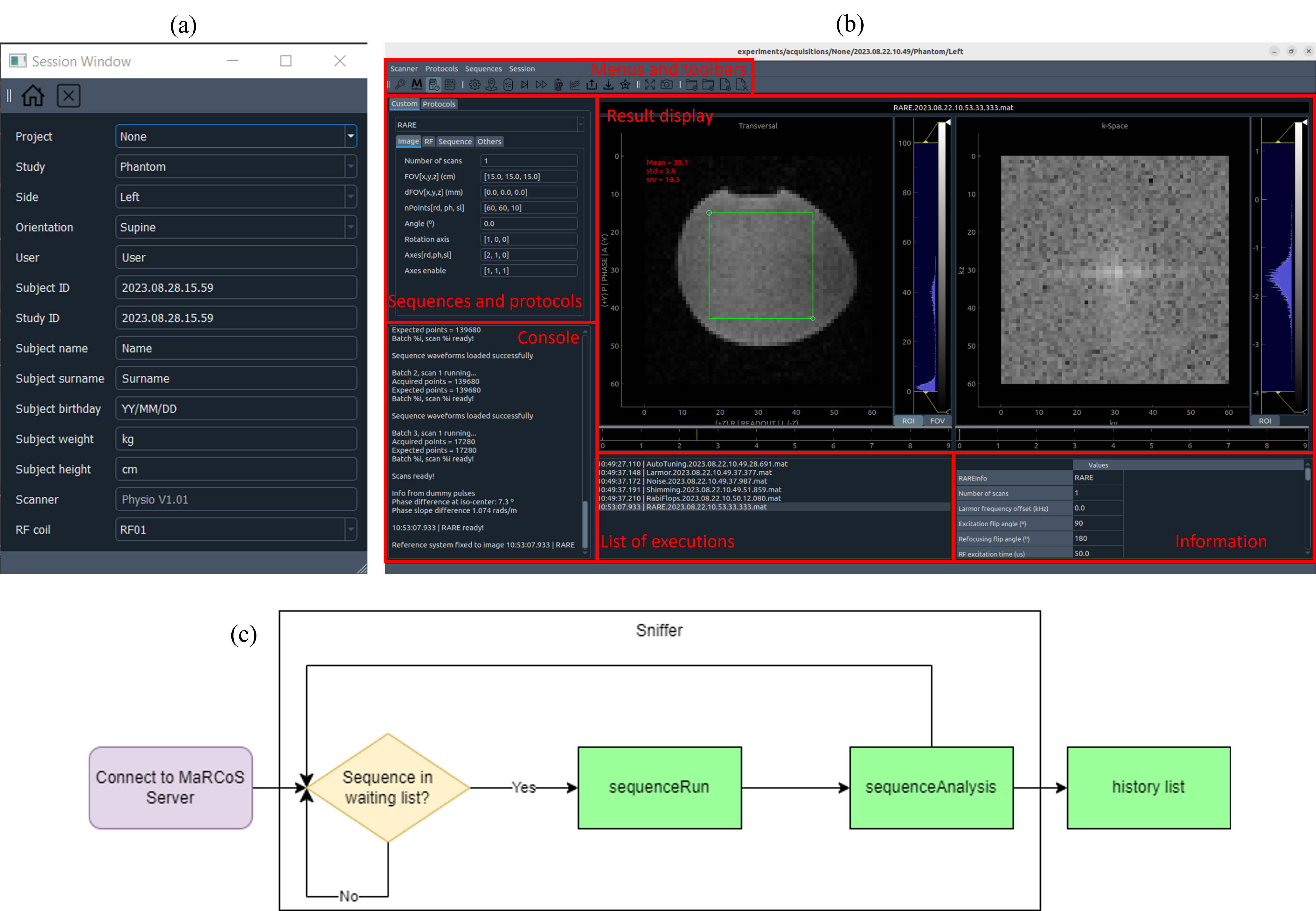}
	\caption{Screenshot of the session (a) and main (b) windows of the \MG{} GUI. (c) Workflow for sequence execution.}
	\label{fig:gui_screenshot}
\end{figure*}

This section describes the features and functionalities encapsulated within \MG{}, and Fig.~\ref{fig:gui_screenshot} shows a screenshot of the new GUI. The user interface builds on the underlying API, allowing seamless communication with the \M{} server, sequence initialization, parameter adjustments, and oversight of data acquisition.

\subsection{Session window}

The session window (Fig.~\ref{fig:gui_screenshot}(a)) in \MG{} opens up as the GUI is launched and serves as the starting point for the user. Here, they can input general information related to the imaging session, such as the research project or clinical trial, technical aspects about the scanner and protocols, and patient details. Through this interface, operators can ensure that images are linked to accurate metadata, enabling identification, tracking, and organization of imaging data. The session window also contributes to streamlining the workflow, as users can predefine session-specific parameters that are then automatically integrated into sequence setups and image acquisitions. All in all, the integration of user, patient, and project information within the \MG{} GUI is designed to promote efficient data management, facilitate traceability, and support adherence to established medical imaging protocols. Additionally, the information provided in the session window is incorporated into the resulting DICOM files, ensuring that imaging data retains its context and relevance for research or clinical purposes.

\subsection{Main window}

The main GUI window is shown in Fig.~\ref{fig:gui_screenshot}(b). Here, users can fine-tune sequence parameters, optimize timings, and personalize pulse patterns for their imaging needs. The GUI also allows protocol design from scratch, enabling researchers and clinicians to experiment with different sequence configurations. A history list provides access to previous reconstructions and results, to compare data and facilitate informed decision-making.

Integral to the \MG{} GUI is a dedicated area designed to display images and results processed by the system. In essence, this area is meant to become a platform for quality assessment, data analysis, and deciding whether a re-scan or a different sequence is required. To this end, users can zoom and crop the field of view, evaluate local and global signal-to-noise ratios automatically and gauge contrast levels.

The \MG{} GUI offers also a real-time console that serves as an informative dashboard. This presents users with relevant feedback, including system status, progress updates, and any potential alerts. Furthermore, an integrated box offers insights into items within the history list. A toolbar complements the GUI, providing dedicated functionalities to manage communication with the \M{} server, oversee sequence operations, and initiate new protocol creations.

Upon opening the GUI, a persistent ``sniffer'' method starts monitoring for sequences that pend execution. The sniffer remains active as long as the main window is open. Once a sequence is detected in the queue, the sniffer calls the \texttt{sequenceRun} and \texttt{sequenceAnalysis} methods, and the resulting data is saved when the sequence terminates. A conceptual representation of this workflow is depicted in Fig.~\ref{fig:gui_screenshot}(c).

\subsection{Post-processing window}
The post-processing window (Fig.~\ref{fig:DentalI}(f)) has a similar appearance to the main window, but it is intended for analyzing and refining the acquired data. Users can access the post-processing window either through the corresponding button located in the toolbar of the main window or by right-clicking on an item in the list of executions, which opens a context menu (see Fig.~\ref{fig:gui_screenshot}(b)). Here users can apply a range of image processing techniques both in $k$-space and in image space. The available tools are still few and basic, but we expect it to become richer as the number of \M{} and \MG{} users increases. For $k$-space we have zero-padding and cosine-bell filters, whereas images can be denoised with Block-Matching 4D (BM4D) \cite{Maggioni2013} or Gaussian filters. The software can be expanded also with \emph{ad hoc} reconstruction algorithms. By default, \MG{} includes standard Fast Fourier Transform (FFT), Partial Fourier by zero padding or projection onto convex sets (POCS) \cite{Haacke1991}, and Algebraic Reconstruction Techniques (ART, \cite{Kaczmarz1937, GORDON1970471, Gower2015, Galve2020}). All iterative methods have been programmed to be compatible with GPU execution.

\section{Methods}\label{sec:Methods}
For this work, \MG{} was installed in two distinct low-field MRI setups: a laboratory-based scanner for \emph{ex vivo} dental imaging \cite{Algarin2020,Borreguero2023}, and a portable extremity scanner \cite{GuallartNaval2022,Algarin2023} located in the experimental radiology department of a large hospital facility. 

To operate \MG{} for the first time, users are advised to follow the setup procedure detailed in Ref.~\cite{MARGA}. The initial step is the installation of the necessary software dependencies, including \M{} and Python packages. \MG{} can be then launched. In our case, the control computers run on Windows for the dental system and Linux for the portable system, the hardware configuration files were adjusted for each scanner, and the essential Python modules (such as PyQt5, pyqtgraph and others) were installed. 

The Main Window has a button for running a set of automatic calibration sequences with a single click. This procedure includes the determination of: i) the Larmor working frequency; ii) the RF coil efficiency, which is then used to calibrate the amplitude of resonant pulses in a dimensionless scale between 0 and 1 to achieve a given flip angle for a given pulse duration; and iii) the gradient currents that actively shim the main magnetic field, estimated from the full-width-half-maximum (FWHM) of the spectrum of a spin-echo.

In the dental scanner we typically start by manually impedance-matching the coil. In contrast, the RF chain of the extremity scanner counts with a system to impedance-match the coil autonomously to \SI{50}{\ohm} at the working frequency. This process can be triggered from \MG{} using a specialized method from the API, but it is also accessible from the main GUI as any other regular pulse sequence. Furthermore, this method is automatically integrated into the standard calibration process if the hardware for autonomous impedance matching is detected by \MG{}.

In the laboratory-based dental scanner \cite{Algarin2020,Borreguero2023}, exclusively dedicated to \emph{ex vivo} imaging, predefined imaging protocols are not a prerequisite. As a sample for these tests we used a cylindrical phantom with copper sulfate diluted in water with a concentration of 1~\%. Here, the imaging process starts with manual impedance matching of the RF coil, followed by an initial calibration and, finally, the execution of customized pulse sequences. Depending on the application, we typically run Gradient Echo (GRE), Rapid Acquisition with Relaxation Enhancement (RARE), and Pointwise Encoding Time Reduction with Radial Acquisition (PETRA), which are all configured using the custom tab in the main GUI.

On the other hand, the portable scanner \cite{GuallartNaval2022,Algarin2023}, operates in a clinical environment for \emph{in vivo} extremity imaging and thus benefits from the possibility of executing imaging protocols. For instance, we created a specific protocol for knee imaging encompassing two localizers (RARE projection images in transversal and coronal orientation) and four RARE sequences with different contrasts: coronal inversion recovery, sagittal T\textsubscript{1}, sagittal T\textsubscript{2}, and transversal T\textsubscript{1}. The complete process therefore involves an initial calibration to adapt to the new patient or body part, the execution of the localizers to determine the field of view and angulation, and the subsequent execution of the corresponding imaging protocol.

\section{Results}\label{sec:Results}

The following subsections evaluate specifically the system performance and the impact of the new developments on research and clinical protocols. These two use cases are arguably representative of the varied environments where low-field MRI is expected to play an increasingly important role. 

\subsection{Laboratory-based dental MRI scanner}

\begin{figure*}
	\centering
	\includegraphics[width=\textwidth]{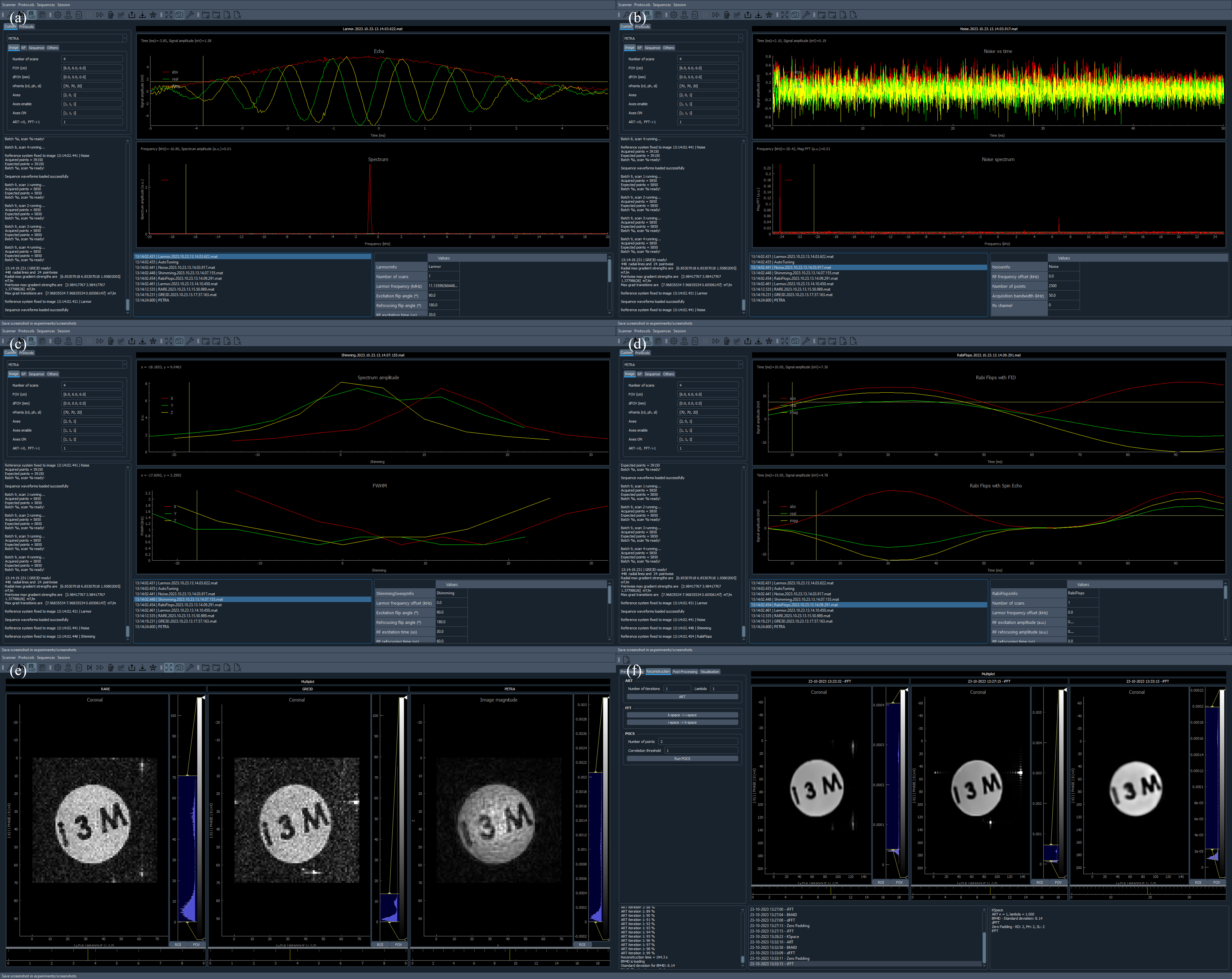}
	\caption{Screenshots of the GUI in the laboratory-based dental MRI scanner: (a) echo to calibrate the Larmor frequency, (b) noise measurement, (c) shimming to optimize the B\textsubscript{0} homogeneity, (d) Rabi flops to calibrate the B\textsubscript{1} efficiency, (e) and (f) from left to right, SE, GRE and PETRA images before and after post-processing.}
	\label{fig:DentalI}
\end{figure*}

Figure~\ref{fig:DentalI} depicts the calibration process in the laboratory-based dental scanner, showcasing various results required for proper system performance: panel (a) shows a spin echo and its corresponding spectrum, required to measure the Larmor frequency, which is at 11.1352~MHz in this example; panel (b) shows a noise measurement and its spectrum, yielding an RMS noise value of \SI{319}{\micro V} at a bandwidth of 50~kHz and revealing distinct electromagnetic interference (EMI) at -24 and 7~kHz relative to the Larmor frequency (these are known to originate at the gradient power amplifier oscillator module); panel (c) displays the signals detected for different gradient configurations during an active shimming calibration, along with their spectral amplitude (top) and full width at half maximum (FWHM, bottom), with an optimal FWHM of 500~Hz (45 ppm); and panel (d) shows Rabi flops obtained by sweeping the excitation time of a single RF pulse, measured through a Free Induction Decay (FID, top) and an RF echo (bottom), where the maximum intensity ($\pi/2$ pulse) is achieved for \SI{30}{\micro s} with an input RF amplitude of 0.34 (arbitrary units between 0 and 1). Following these calibrations, the system is considered prepared for imaging.

In Fig.~\ref{fig:DentalI}(e) we show images reconstructed from the signal detected during the execution of three distinct pulse sequences: RARE (left), GRE (center), and PETRA (right). The post-processed versions of these images are displayed in panel (f). Specifically, we applied a BM4D filter for denoising and rescaled the matrix to double its size.

\subsection{Hospital-based MRI Scanner for Extremity Imaging}
\begin{figure*}[t]
	\centering
	\includegraphics[width=\textwidth]{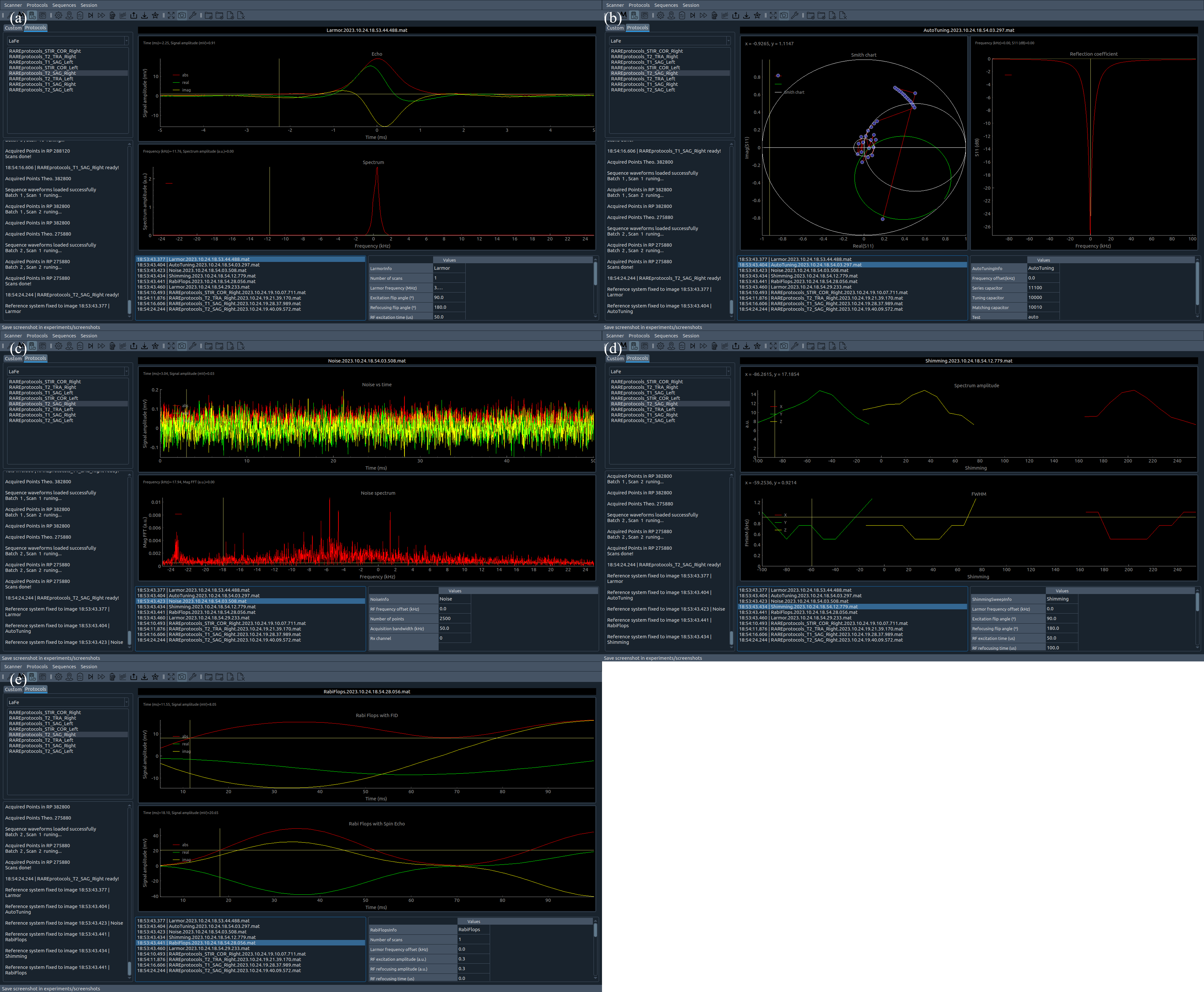}
	\caption{Screenshots of the GUI in the hospital-based MRI scanner for extremity imaging: (a) echo to calibrate the Larmor frequency, (b) automatic impedance tuning and matching; (c) noise measurement, (d) shimming to optimize the B\textsubscript{0} homogeneity, (e) Rabi flops to calibrate the B\textsubscript{1} efficiency.}
	\label{fig:PhysioI}
\end{figure*}

Figure~\ref{fig:PhysioI} provides an overview of the calibration process in the hospital-based extremity MRI scanner: panel (a) shows the spin echo and its corresponding spectrum for the determination of the Larmor frequency at 3.0673~MHz; panel (b) displays the automatic impedance RF coil matching procedure in the Smith chart (left) and the resulting reflection coefficient after the impedance matching (right), reaching -27~dB with a bandwidth of 20~kHz, giving a Q factor $\approx150$; panel (c) shows the noise measurement and its spectrum, yielding an RMS noise value of \SI{58}{\micro V} at 50~kHz, twice the expected Johnson noise level; panel (d) shows the shimming calibration curves, with an optimal value of FHWM $\approx0.5$~kHz (163~ppm); and panel (e) shows the Rabi flops, providing a maximum signal for an RF excitation time of \SI{35}{\micro s} for an input RF amplitude of 0.30 a.u.

\begin{figure*}[t]
	\centering
	\includegraphics[width=\textwidth]{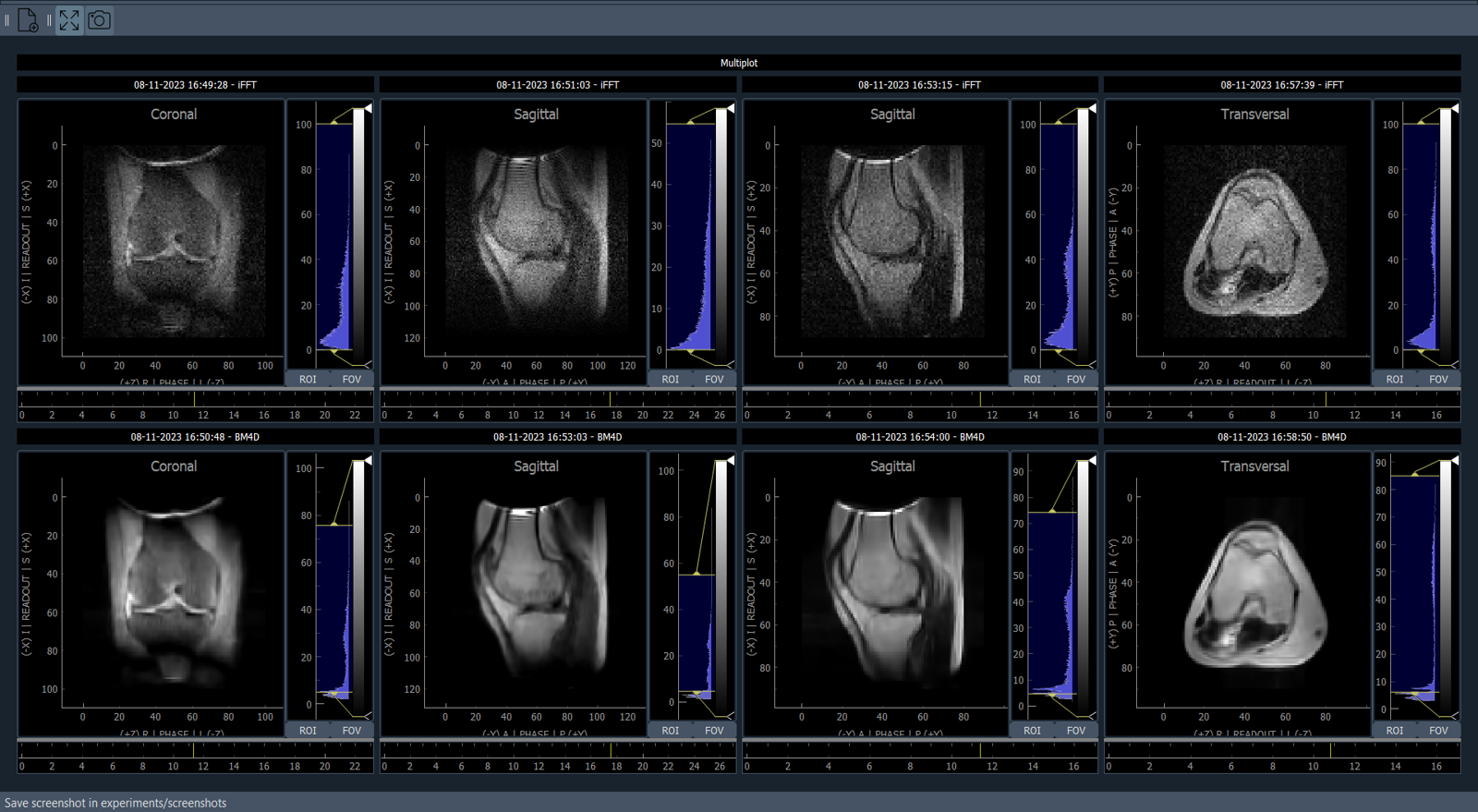}
	\caption{Images resulting from the clinical protocol executed on the extremity scanner. From left to right: coronal inversion recovery, sagittal T\textsubscript{1}, sagittal T\textsubscript{2}, and transversal T\textsubscript{1}. Top (bottom) shows the raw (post-processed) images.}
	\label{fig:PhysioII}
\end{figure*}

Following calibrations, we present the acquired and post-processed images in Fig.~\ref{fig:PhysioII}. From left to right, the four images correspond to coronal inversion recovery, sagittal T\textsubscript{1}, sagittal T\textsubscript{2}, and transversal T\textsubscript{1}. The top images are raw and post-processing in the bottom included only BM4D filtering, except for sagittal T\textsubscript{1}, where a cos-bell filter was applied in the $k$-space readout direction to suppress artifacts observed in the raw image.

\section{Discussion}\label{sec:Discussion}

The main qualitative conclusion we extract from operating with \MG{} is that the integration between GUI and API provides users with simplified but sufficient control over vastly different MRI scanners, as demonstrated in the above use cases. Notably, multi-sequence imaging protocols have been a crucial development for the hospital-based system for routine \emph{in vivo} scans, and the possibility to fine-tune imaging parameters has enabled tailored sequences and, ultimately, an improvement in image quality also in our laboratory-based scanner, where we only worked with phantoms. Additionally, the one-click calibration process simplified significantly the optimization of the scanners' performance, reducing time and manual errors that often compromise the diagnostic potential of our images as well as the patient experience. As an illustrative example of the software usability, it took a newly arrived student around two hours to operate autonomously the clinical scanner. In summary, the enhancements aimed at improving user interaction with \M{} have been implemented and have led to a more intuitive experience and, perhaps most importantly, \M{} is now usable within medical environments.

\section{Conclusion}\label{sec:Concl}

The new \MG{} GUI has been designed to emerge as a transformative element in the landscape of open-source MRI control, redefining user interaction and amplifying the potential for advanced medical imaging. This paper has explored the multifaceted dimensions of the software, presenting its features and functionalities.

An overarching goal of \MG{} is to contribute to the democratization of MRI control and, in that way, to MRI access. The GUI's interactive components, including the image and results display area, is designed to allow users to validate data in real-time, make informed decisions, and collaborate effectively with colleagues.

Looking ahead, we plan to make \MG{} compatible with Pulseq, a widely embraced open-source framework written in Matlab and under development for Python that is extensively used within the MRI community for programming pulse sequences \cite{Layton2017}. Compatibility between \M{} and Pulseq has been already demonstrated \cite{GuallartNaval2022b}, and transferring this to \MG{} should foster its early adoption by a larger community. Similarly, we plan to expand \MG{} to save and operate with raw data in the ISMRMRD standard format \cite{Inati2017}, which was designed to encapsulate complex MRI data comprehensively.  By adopting ISMRMRD, we shall enhance interoperability across MRI systems and software platforms.

Finally, official regulatory approval by regional governing bodies will eventually be required for the use of \M{} and \MG{} within healthcare systems. This is not our particular priority at the moment, but our developments are indeed influenced by the European Medical Device Regulation (Regulation (EU) 2017/745 of the European Parliament and of the Council) and United States Food \& Drug Administration directives for medical devices (510(k), since the device is low to moderate risk and there is a legally marketed predicate), and we are making a conscious effort to adhere to the clinical standards defined therein.

\section*{Author contributions}
Software programmed by JMA, TGN and JB. Low-field images taken by TGN, JB and JMA. Data analysis and evaluation by all authors. Project conceived and supervised by JMA, FG and JA. Paper written by JMA and JA, with input from all authors.

\section*{Acknowledgment}
We thank Luis Mart\'i-Bonmat\'i, Amadeo Ten-Esteve and Sonia Gin\'es-C\'ardenas from IIS La Fe for their clinical collaboration, Dhareena Comlan for contributions to the post-processing \MG{} platform, and the open \M{} community for their continued support to the project. Project funded by the European Innovation Council (grant number 101136407), Ministerio de Ciencia e Innovaci\'on (PID2022-142719OB-C22), Agencia Valenciana de la Innovaci\'on (INNVA1/2022/4 and INNVA1/2023/30), and EURAMET (22HLT02).

\section*{Ethical statement}
All participants in this work were adults and provided written informed consent for this study. Ethical approval was obtained from the Ethics Committee (CEIm) of La Fe Hospital in Valencia (CEIm-F-PE-01-16, research agreement number 2022-187-1).

\section*{Conflict of interest}
No competing interests.


\end{document}